\def\pr#1{ Phys.Rev. {\bf B#1}}

\def\.#1{\mathaccent 95#1}
\def\^#1{\mathaccent 94 #1}
\def\~#1{\mathaccent "7E #1}

\def\eq{\enskip =\enskip}
\def\pls{\enskip +\enskip}
\def\mns{\enskip -\enskip}

\def\H{{\bf H}}
\def\I{{\bf I}}
\def\Pr{{\bf P}}
\def\Tr{{\bf T}}

\documentstyle[12pt]{article}
\begin{document}
\baselineskip 24pt
\title{\bf AN AUGMENTED SPACE RECURSION STUDY OF THE ELECTRONIC STRUCTURE OF ROUGH
EPITAXIAL OVERLAYERS }
\author{
{\bf Biplab Sanyal, Parthapratim Biswas, A. Mookerjee}\\
{\normalsize S.N. Bose National Centre for Basic Sciences}\\
{\normalsize JD Block, Sector 3, Salt Lake City, Calcutta 700091, INDIA}\\
{\normalsize }\\
{\bf Hemant G Salunke and G.P. Das}\\
{\normalsize Bhabha Atomic Research Centre, Trombay, Mumbai , INDIA}\\
{\normalsize and}\\
{\bf A.K. Bhattacharyya}\\
{\normalsize Centre for Catalysis and Materials Studies,}\\
{\normalsize Department of Engineering, University of
Warwick, Coventry, England}\\
}
\newpage
\maketitle
\begin{abstract}
In this communication we propose the use of the Augmented Space 
Recursion as an ideal methodology for the study of electronic and
magnetic structures of rough surfaces, interfaces and overlayers.
The method can take into account roughness, short-ranged clustering
effects, surface dilatation and interdiffusion. We illustrate our
method by an application of Fe overlayer on Ag (100) surface.
\end{abstract}
\newpage

\section{Introduction}

Magnetism at surfaces, overlayers and interfaces has evoked much interest
in recent times \cite{kn:feder}. The chemical environment of an atom at
a surface or overlayer is very different from the bulk. The difference  in
environment, existence of surface states and hybridization of the states
of the overlayer with those of the substrate can give rise to a wide variety
of new and interesting material and magnetic properties. This  wide variety has the potential
for being the basis of surface  materials design. This is the underlying reason for   the
absorbing theoretical interest in this field.

In this communication we wish to argue that the Augmented Space Recursion (ASR) introduced
by us earlier is one of the most suitable techniques for the study of rough overlayers
and interfaces.

First principles  all electron  techniques for  the determination of the electronic
structure based  on the local spin density approximation (LSDA) have made reasonably
accurate quantitative calculations possible. Originally the most  popular of the methods
was the parametrized tight-binding or the linear combination of atomic orbitals (LCAO)
method \cite{kn:lcao}. However, the fact that the parametrized hamiltonian is, in general,
never transferable and that the basis does not have sufficient variational freedom, has
led to the eclipse of such methods for quantitative calculations ; in particular of
properties as sensitive to these assumptions as the magnetic moment . There have been attempts
of resuscitating the LCAO by introducing  ideas of environment dependent parametrization \cite{kn:fab}
. The generally accepted quantitative techniques include the Augmented Plane Wave (APW) and its
linearized version (LAPW)\cite{kn:lapw} and the Korringa-Kohn-Rostocker (KKR) and its linearized
version (LMTO)\cite{kn:lmto}. The two basically related methods come both  in the Full Potential versions
where no assumption is made about the shape of the charge density or the potential, or in the
spherically symmetrized Muffin-tin Potential versions. The electrons may be treated either
semi-relativistically or fully relativistically  \cite{kn:rkkr}. In addition, Andersen
and co-workers \cite{kn:tblmto} have proposed a tight-binding LMTO (TB-LMTO) where the 
real space representation of the hamiltonian is sparse. Which of the two basic methods we
choose often depends on a matter of taste and history. Moreover, how far we wish to go down
the ladder of different approximations is guided by the accuracy required and the computational heaviness
we wish to face. We would not like to comment on this, other can justifying the specific
technique we have chosen for ourselves.

The other important aspect of the problem is the loss of translational symmetry perpendicular to the
surface. This aspect has been  dealt with by different authors in different ways : 
\begin{enumerate}
\item[(i)] finite slab calculations, which assume
that finite size effects are negligible  \cite{kn:finite}
\item[(ii)]  supercell calculations, where the
translational symmetry is restored. Each supercell has a replica  of the finite system and
the assumption is that the supercells are large enough so as not to affect one another 
\item[(iii)] the
slab Green function method where the translational symmetry parallel to the surface is utilized
and the perpendicular direction is treated in real space \cite{kn:kd1}-\cite{kn:g}
. The embedding
method of Inglesfield and coworkers \cite{kn:embed} belongs to this group, where the Green
function of the semi-infinite solid is calculated by downfolding onto this semi-infinite subspace.
\item[(iv)] the fully real space based Recursion method \cite{kn:vol35} which does not require
any translational symmetry and was originally developed for dealing with surfaces and interfaces.
\end{enumerate}

 Overlayers produced by 
molecular beam epitaxy and other vapor deposition techniques are, by and large,
rough. Local probes, such as STM techniques, reveal steps, islands and pyramid-like
structures.  Moreover, there is always interdiffusion between the overlayer and the
substrate leading to a disordered alloy like layer at the interface. This brings in the last
important aspect of the problem : roughness or disorder parallel to the surface. A majority
of the theoretical work done on surfaces and overlayers so far had always assumed flat
layers. 
 These generally involve the use of surface Green functions,
G(k$_{\parallel}$,z), which allow breaking of translational symmetry perpendicular to the
surface, but  presume such symmetry parallel to  it \cite{kn:kd1, kn:kd2}. 
Roughness has been introduced in overlayers by randomly alloying it with {\sl empty spheres}
 \cite{kn:g}. Such alloying has been assumed to be homogeneous and has been treated within 
a mean field or the coherent potential approximation (CPA) . Attempts at going beyond
the CPA has not been generally successful. One of the more successful approaches in this
direction is the Augmented Space Formalism (ASF)\cite{kn:asf}  and techniques basically based on it,
like the travelling cluster approximation (TCA) \cite{kn:tca}.

Let us now justify why we wish to introduce the Augmented Space Recursion based on the TB-LMTO as
an attractive method for the study of rough surfaces, overlayers or interfaces.

The CPA has proven to be an accurate approximation in a very large body of applications. Why then do
we wish to go beyond ? We should recall that the CPA is {\sl exact} when the local coordination is
infinite. Its accuracy is inversely proportional to the local coordination. We therefore expect
the CPA to be comparatively less accurate at a surface as compared with the bulk  calculations. 
Further, the CPA basically describes homogeneous randomness. It cannot accurately take into account
clustering, short-ranged ordering or local lattice distortions, of the kind we expect to encounter
in the  rough surfaces produced experimentally. The ASF allows  us to describe exactly such
situations, without violating the so called ``herglotz" properties which the  approximated
averaged Green function must possess \cite{kn:asf2}.

We shall combine the ASF with the Recursion method to calculate the configuration averaged Green
functions. We should note that the Augmented Space Theorem is {\sl exact} \cite{kn:asf2} and the
approximation involves in terminating the recursion-generated continued fraction. Analyticity preserving
``terminators" have been introduced by Haydock and Nex \cite{kn:hn} and Lucini and Nex \cite{kn:ln}.
Recently Ghosh {\sl et.al.} \cite{kn:gdm1} have discussed the convergence of the 
Augmented Space Recursion and indicated
how to generate physical quantities within a prescribed error window. The Recursion method, being entirely
in real space, does not require any translational symmetry and is ideally suited for systems with
inhomogeneous disorder.  However, for the Recursion method to be a practicable computational technique,
we must choose a basis of representation in which the effective hamiltonian is sparse, i.e. short
ranged in real space. The best choice of a computationally simple yet accurate basis is the TB-LMTO.
This is what we describe in this communication. 
However, the screened-KKR \cite{kn:rkkr} would also be a more quantitatively accurate 
 choice. We would require the energy
dependent extension of the Recursion  method. This has been developed recently \cite{kn:gdm2} and its
application to the screened-KKR will be described in a subsequent communication.

To illustrate the method we shall take a well studied example : that of 
 Fe deposited on the (100) surface of a Ag substrate.
The lattice parameter of bcc Fe, the most commonly known ferromagnet \cite{kn:hc}
matches the nearest neighbour distance on the (100)  surface of fcc Ag (half the face
diagonal), a very good  
non  magnetic  electrical  conductor.  This favours epitaxial deposition of
bcc Fe on Ag(100) manifesting  interesting  magnetic  properties.  

Before describing the methodology in some detail we need to clarify the following
point : in order to describe inhomogeneous disorder we have taken recourse to
the Generalized Augmented Space Theorem \cite{kn:mp}.
 This generalized  ASF does take into account short-ranged order through
the Warren-Cowley parameter and yields an analytic  herglotz  approximation. In a
recent publication \cite{kn:gk} the authors make the strange statement that the
generalized ASF yields negative densities of states and quotes the work of Razee and
Prasad \cite{kn:rp}. The statement is untrue and the misconception should be cleared up.
A careful reading of the quoted article \cite{kn:rp} will show that in applying the
generalized ASF Razee and Prasad use the Nikodym-Radon transform and write the joint
density of states of the hamiltonian parameters ${\cal P}(\{ \epsilon_{i}\})$ 
 as $\left( \prod p(\epsilon_{i})\right) \Phi(\{\epsilon_{i}\})$ . For homogeneous
disorder $\Phi(\{\epsilon_{i}\})$ is unity, while for inhomogeneous disorder the
authors expand the function as an infinite series involving various correlation
functions between the $\{\epsilon_{i}\}$ ( the simplest two site correlation can
be written in terms of the Warren-Cowley parameter). They then truncate this series
after a few terms. This extra approximation cannot guarantee the preservation
of the herglotz analytic properties and is the cause of the observed negative
density of states in some energy regimes. The generalized ASF described by Mookerjee
and Prasad \cite{kn:mp} does not take recourse to such an approximation and has been
shown to be exact in the referenced paper. Approximation then arises entirely due to the
recursion termination - which has been shown to preserve the herglotz analytic properties.

\section{The Generalized Augmented Space Theorem}

In this section we shall describe the generalized Augmented Space Formalism.
The hamiltonian is a function of a set of random variables $\{ n_{i}\}$ which
are not independent, so that the joint probability distribution can be written 
in terms of the conditional probability densities of the individual variables as :

\[ p(\{n_{i}\}) \eq p(n_{1}) \; \prod_{k} p\left( n_{k}\vert n_{k-1},n_{k-2},
\ldots n_{1}\right)\]

Each random variable $n_{k}$ has associated with it its own configuration space
$\Phi_{k}$ and, in the case of correlated disorder, a set of operators
$\{ M_{k}^{\lambda_{k-1},\lambda_{k-2},\ldots \lambda_{1}}\}$  whose spectral
density are the conditional probability densities of the random variable,
dependent
on the configurations of the previous labeled ones. The $\lambda_{k}$
label the configurations of the variable $n_{k}$. The configuration
space of the set of random variables is the product $\Psi$ \eq $\prod^{\otimes}
_{k} \Phi_{k} $. 
 What the generalized
augmented space theorem proved was that, if we define operators on this
full configuration space,

\[ \tilde{M}_{k} \eq \sum_{\lambda_{1}}\sum_{\lambda_{2}}\ldots\sum_{\lambda_{k-1}}
\; P_{1}^{\lambda_{1}}\otimes P_{2}^{\lambda_{2}}\otimes \ldots
P_{k-1}^{\lambda_{k-1}} \otimes M^{\lambda_{k-1},\lambda_{k-2},\ldots \lambda_{1}}
_{k} \otimes I \otimes I \ldots \]

then the configuration average of any function of the hamiltonian is given {\sl exactly}
by :

\begin{equation}
<< {\cal{F}}(\{n_{k}\})>> \eq \langle F^{0} \vert \tilde{{\cal{F}}}\left( \{
\tilde{M}_{k}\} \right) \vert F^{0}\rangle
\end{equation}

The average state $\vert F^{0}\rangle$ is defined by :

\begin{eqnarray*}
\vert F^{0}\rangle & \eq  & \prod_{k} \vert f^{0}_{k} \rangle \\
\vert f^{0}_{k} \rangle &\eq & \sum_{\lambda_{k}} \sqrt{\omega_{\lambda_{k}}
^{\lambda_{1},\lambda_{2}\ldots\lambda_{k-1}}} \vert \lambda_{k}\rangle\\
\end{eqnarray*}

where the numbers under the root sign are the conditional probability weights
for the various configurations of the variable $n_{k}$. 

In our model, the random variables are the occupation variables of a site by
two different kind of atoms . The simplest model is one that  assumes that
the occupation of the nearest neighbours of a site depends on its own occupation. The probability densities
are given by :

\begin{eqnarray*}
p(n_{1}) &\eq & x\;\delta(n_{1}-1)\pls \; y\;\delta(n_{1}) \\
p(n_{2}\vert n_{1}=1) & \eq & (x+\alpha y)\;\delta(n_{2}-1)\pls (1-\alpha)y\;\delta(n_{2}) \\
p(n_{2}\vert n_{1}=0) &\eq & (1-\alpha)x\;\delta(n_{2}-1)\pls (y+\alpha x)\;\delta(n_{2}) \\
\end{eqnarray*}

Where x and y are the concentrations of the constituents and  $\alpha$
is the Warren-Cowley short-ranged order parameter. $\alpha$=0 refers to the completely
random case, when the various operators $M_{k}^{\lambda_{k-1},\ldots \lambda_{1}}$
become independent of the superscripts and the generalized augmented space theorem
reduced to the usual augmented space theorem. $\alpha\;<\; 0$ indicates tendency towards ordering
alternately, while $\alpha\; >\; 0$ indicates tendency towards segregation. 

The representations of the corresponding operators required are the following :

\[
M_{1} \eq \left(  \begin{tabular}{cc}
                   x & $\sqrt{xy} $\\
                   $\sqrt{xy}$ & y \\
                  \end{tabular} \right) \]

\[
M_{2}^{1} \eq \left( \begin{tabular}{cc}
                     x+$\alpha$y & $\sqrt{(1-\alpha)y(x+\alpha y)}$ \\
                      $\sqrt{(1-\alpha)y(x+\alpha y)}$ & (1-$\alpha$)y  \\
                    \end{tabular} \right) \]

\[
M_{2}^{0} \eq \left( \begin{tabular}{cc}
                     (1-$\alpha$)x &  $\sqrt{(1-\alpha)x(y+\alpha x)}$ \\
                      $\sqrt{(1-\alpha)x(y+\alpha x)}$ & y+$\alpha $x \\
                    \end{tabular} \right) \]

\[
P_{1}^{0} \eq \left( \begin{tabular}{cc}
                     x & $\sqrt{xy}$ \\
                   $\sqrt{xy}$ & y \\
                    \end{tabular} \right) \]

\[
P_{1}^{1} \eq \left( \begin{tabular}{cc}
                     y & $-\sqrt{xy}$ \\
                   $-\sqrt{xy}$ & x \\
                    \end{tabular} \right) \]

\section{TB-LMTO-ASR formulation}

Our system consists of  a semi-infinite Ag substrate
with layers of Fe atoms on the (100) surface. 
 We shall describe the hamiltonian of the electrons within a tight-
binding linearized muffin-tin orbitals basis (TB-LMTO). As described
earlier, we shall take care of the charge leakage into the vacuum by 
layers of empty spheres containing charge but no atoms. We shall roughen
the topmost layer by randomly alloying the Fe atoms with empty spheres. We shall
allow for short-ranged order in the alloying. Segregation will imply that the
Fe atoms and empty spheres cluster together forming islands and clumps.
Ordering on the other hand will imply that Fe atoms like to be surrounded
by empty spheres and vice versa.

The details of the description of the effective augmented space hamiltonian
 has been described at length in an earlier paper \cite{kn:sbm}. We shall
indicate the generalization of that result when nearest neighbour short-ranged
order is introduced as described above.

\begin{eqnarray}
\tilde{H} &\eq & \H_{1} \tilde{\I} \pls \H_{2} \sum_{k}
\Pr_{k}\otimes\Pr^{k}_{\downarrow} \pls \H_{3}
\sum_{k}\Pr_{k}\otimes\{\Tr^{k}_{\downarrow\uparrow}
+\Tr^{k}_{\uparrow\downarrow}\}\nonumber\\
& &  \pls \H_{4}\sum_{k}\sum_{k'}\Tr_{kk'}\otimes \I\pls 
 \pls \alpha\H_{2}\sum_{m\epsilon N_{1}} \Pr_{m}\otimes\Pr^{1}_{\downarrow}
\otimes \{ \Pr^{m}_{\uparrow}- \Pr^{m}_{\downarrow} \} \pls
\nonumber\\ 
& & \pls \H_{5}\sum_{m\epsilon N_{1}} \Pr_{m}\otimes \Pr^{1}_{\uparrow}\otimes \{\Tr^{m}_
{\uparrow\downarrow}+\Tr^{m}_{\downarrow\uparrow}\}\pls 
 \pls\H_{6} \sum_{m\epsilon N_{1}} \Pr_{m}\otimes \Pr^{1}_{\downarrow}\otimes
\{ \Tr^{m}_{\uparrow\downarrow}+\Tr^{m}_{\downarrow\uparrow}\} \pls\nonumber\\
& & \pls\alpha\H_{2} \sum_{m\epsilon N_{1}} \Pr_{m}\otimes\{\Tr^{1}_{\uparrow\downarrow}+\Tr^{1}_{\downarrow\uparrow}\}
\otimes \{ \Pr^{m}_{\uparrow}-\Pr^{m}_{\downarrow}\}\pls\nonumber\\
& &  \pls 
\H_{7} \sum_{m\epsilon N_{1}} \Pr_{m}\otimes\{\Tr^{1}_{\uparrow\downarrow}+\Tr^{1}_{\downarrow\uparrow}\}
\otimes\{\Tr^{2}_{\uparrow\downarrow}+\Tr^{2}_{\downarrow\uparrow}\}\nonumber\\
\end{eqnarray}

where, $N_{1}$ are the set of nearest neighbours of the site labeled 1 on
the surface and for calculations of averaged local densities of states at a constituent
labeled by $\lambda$ we have \\

\begin{eqnarray}
 \H_{1} & \eq &A(C/\Delta)\Delta_{\lambda} \mns \left( E\;A(1/\Delta)\Delta_{\lambda}
- 1\right) \nonumber\\
\H_{2} &\eq & B(C/\Delta)\Delta_{\lambda} \mns  E\;B(1/\Delta)\Delta_{\lambda}\nonumber\\
\H_{3} & \eq & F(C/\Delta)\Delta_{\lambda} \mns E\; F(1/\Delta)\Delta_{\lambda}\nonumber\\
\H_{4} & \eq & \left(\Delta_{\lambda}\right)^{-1/2} S_{RR'} \left(\Delta_{\lambda}\right)^{-1/2}\nonumber\\ 
\H_{5} & \eq & F(C/\Delta)\Delta_{\lambda} \left[ \sqrt{(1-\alpha)x(x+\alpha y)}
+\sqrt{(1-\alpha)y(y+\alpha x)} -1 \right] \nonumber\\
\H_{6} & \eq & F(C/\Delta)\Delta_{\lambda} \left[y\sqrt{(1-\alpha)(x+\alpha y)/x}+
x\sqrt{(1-\alpha)(y+\alpha x)/y}-1\right] \nonumber\\
\H_{7} & \eq & F(C\Delta)\Delta_{\lambda} \left[ \sqrt{(1-\alpha)y(x+\alpha y)}
-\sqrt{(1-\alpha)x(y+\alpha x)} \right]\nonumber\\
\end{eqnarray}

\begin{eqnarray*}
A(Z) & \eq & x\; Z_{A} \pls y\; Z_{B} \nonumber\\ 
B(Z) & \eq & (y-x) \; \left( Z_{A}-Z_{B}\right) \nonumber\\
F(Z) & \eq & \sqrt{xy} \; \left( Z_{A}-Z_{B}\right)\nonumber \\
\end{eqnarray*}

The C, $\Delta$ and S are matrices in angular momenta, the first two being
diagonal. We note first of all that when the short-ranged order disappears
and $\alpha$ \eq 0, the terms \H$_{5}$ to \H$_{7}$ also becomes zero and
the hamiltonian reduces to the standard one described earlier \cite{kn:sbm}.

This effective hamiltonian is sparse in the TB-LMTO basis, but as the
expressions show there is an energy dependence in the first three terms.
This compels us to carry out recursion at every energy step. However, Ghosh
et. al. \cite{kn:gdm2} have shown that the corresponding energy dependence of
the continued fraction coefficients is very weak and if we carry out 
recursions at a few selected {\sl seed} energies across the spectrum, we
may obtain accurate results by spline fitting the coefficients over the
spectrum.

For the self-consistent calculations we require to calculate the partial (atom projected)
density of states at various sites in different layers. This is done 
by running the recursion starting from sites in different layers.
We shall assume that after 5 layers from the surface bulk values are obtained.
We checked that this is indeed the case, by comparing the results for the 5-th
layer and a full bulk calculation. The Fermi-energy of the system is that of the
bulk substrate which we have taken from the bulk calculations.
 In all cases we have used upto seven shells in augmented space and
terminated the recursion after 8-10 steps of recursion. We have used the terminator
proposed by Lucini and Nex \cite{kn:ln}. As discussed in an earlier paper \cite{kn:gdm1}
, we  have made sure that the moments of the densities of states converges with the
number of augmented space shells and recursions within a preassigned error range, which
is consistent with the errors made in the TB-LMTO approximations. We have made the
recursive calculations LDA self-consistent. For this we had to obtain the radial
solutions of the Sch\"odinger equation involving the spherically symmetric LDA
potential 

\[ V^{\lambda}_{p} (r) \eq -2 \frac{Z^{\lambda}}{r} \pls V^{\lambda,H}_{p}\left[\rho^{\lambda}(r)\right]
\pls V^{\lambda,XC}_{p} \left[ \rho^{\lambda}(r)\right] \pls \sum_{L}\sum_{q} M^{L}_{pq}
Q^{L}_{q} \]

$\lambda$ labels the type of atom, Z$^{\lambda}$ its atomic number, p labels the
particular layer. The second term in the equation is the Hartree potential, which is
obtained by solving the Poisson equation with the layer and atom projected charge
densities. The third term is the exchange-correlation term. For this term we
have used the Barth-Hedin form. In the last term

\[ Q^{L}_{p} \eq \sum_{\lambda} x^{\lambda}_{p} \left\{ \frac{\sqrt{4\pi}}{2\ell +1}
\int_{0}^{s} Y_{L}(\hat{r})\vert r\vert^{\ell} \rho^{\lambda}_{p}(r)dr - Z^{\lambda}\delta_{\ell,0}\right\}\]

Here $\lambda$ for the overlayer is either Fe or empty sphere and the concentrations
x$^{\lambda}_{p}$ is either x or(1-x). For the substrate $\lambda$ refers only to Ag and
its concentration is 1, while for the charge layers outside the overlayer $\lambda$
refers to the empty sphere and its concentration is also 1.

This last term describes the effect of redistribution of charge near the surface which is
particularly important for surface electronic structure. This charge density near the
surface is far from spherically symmetric. We have taken both the monopole ($\ell$=0,m=0)
and the dipole ($\ell$=1,m=0) contributions. We have also averaged the multipole
moments in each layer and used the technique described by Skriver and Rosengaard \cite{kn:sr}
to evaluate the matrices $M^{L}_{pq}$ by an Ewald technique.

\section{Results and Discussion.}

In order to compare our results with calculations carried out earlier, we
shall first carry out calculations on a (100) surface of bcc Fe. Earlier
Wang and Freeman \cite{kn:lcao} had used the LCAO method for the study of the
same system. The FP-LAPW had been used by Ohnishi {\sl et.al.} \cite{kn:lapw}
also the study the (100) surface of bcc Fe. The bulk lattice parameter
was chosen (as in the case of Ohnishi {\sl et.al.}) to be 5.4169 a.u.
At this stage no lattice relaxation was considered. The results quoted
below were for the semi-relativistic self-consistent LSDA TB-LMTO both
supercell and ASR.
The following table  compares the magnetic moment per atom for the three
different methods quoted above :

\begin{center}
\begin{tabular}{|c|c|c|c|c|}\hline
  & S & S-1 & S-2 & B \\ \hline
Wang and Freeman & 3.01 & 1.69 & 2.13 & 2.16 \\
Ohnishi {\sl et.al.}& 2.98 & 2.35 & 2.39 & 2.25 \\
Sanyal {\sl et.al.}$^{(a)}$ & 2.86 & 2.16 & 2.38 & 2.17 \\ 
Sanyal {\sl et.al.}$^{(b)}$ & 2.99 & 2.17 & 2.38 & 2.27 \\ \hline
\end{tabular}
\end{center}
\vskip -0.2cm
\centerline{\small {\bf Table 1} Magnetic moments in bohr-magnetons/atom}
\vskip -0.1cm
\centerline{\small $^{(a)}$ supercell and $^{(b)}$ ASR calculations}

Our central layer magnetic moment per atom is close to the bulk value
given by Wang and Freeman and slightly lower than that given by Ohnishi
{\sl et.al.}. All three methods exhibit Friedel oscillations in the
magnetic moment, although Wang and Freeman's oscillations are larger
than both Ohnishi {\sl et.al.} and our work. Our magnetic moment at the
surface layer is rather small as compared to the earlier works. However,
in these initial calculations (shown as (a) in the table) we have not taken into account surface
relaxation. Local lattice relaxation can be easily taken into account
within the TB-LMTO-ASR  \cite{kn:relax}. We refer the reader to the
details of the relaxation method in the reference mentioned. A 7-8 $\%$
relaxation of the surface layer leads to a surface magnetic moment of
2.99 $\mu_{B} /atom$ which is in good agreement with both the earlier works (shown as (b) in Table 1).

We shall now turn to the study of  Fe (100) on the (100) surface of fcc
Ag substrate. 
We shall carry out the calculations using two different techniques. First,
we  shall use  the  Tight  Binding Linearised Muffin Tin Orbital (TB-LMTO)
method   with  a minimal  (s,p,d)
basis  set  for  Fe  and  Ag  sites  in a  tetragonal supercell. Both spin
polarized as well as non-spin polarized calculations were performed 
on a  Fe/Ag multilayer containing
a monolayer of Fe, a monolayer of empty spheres above them   and four Ag layers
as the substrate. 
The empty  spheres  take care of the charge leakage into the vacuum across
the free surface.
 The results of the calculation show  that  spin  polarization
yields  a  lower  total ground state energy as compared with the  unpolarised case by
$\sim $0.092 eV/atom suggesting that the  ground  state  is  magnetic .
  All the Fe layers have  ferromagnetically arranged 
moments with interface Fe layers having a magnetic moment of $\sim  2.86  \mu
_B$  (bulk  value 2.27 $\mu _B$). Also Fe induces a ferromagnetic moment in
Ag at the interface of $\sim 0.012 \mu _B$  per  atom.  
 The  calculation  also  suggests  Friedel
oscillations in net valence charge in Ag as one goes from interface to bulk
in Ag. 
This is because of moment spillage into the empty spheres. Such moment spillage outside the
surface has also been observed by Ohnishi {\sl et.al. } \cite{kn:lapw}.

We shall refine our calculations in three steps. First we shall introduce the
local lattice relaxation technique within the TB-LMTO-ASR \cite{kn:relax}
to relax the surface layer. We shall inflate the interlayer distance
between the surface layer and the one just below it. Figure 1 shows the
variation of the magnetic moment at the surface layer as a function of
the percentage lattice dilatation at the surface. The minimum of the
total energy occurs at around 7.5$\%$ dilatation. Here the moment carried
by the monolayer of Fe is 3.17 $\mu_{B} /atom$, which is not very far from
the value of 3.1 $\mu_{B} /atom$ quoted by Bl{\" u}gel based on FP-LAPW calculations \cite{kn:bl}.

Next we  shall begin with a  planar monolayer of Fe on Ag  
 and  roughen the monolayer by alloying it with empty spheres. We shall
now use the self-consistent ASR for obtaining the electronic density of states
and local magnetization as a function of the concentration of alloying and the
short-range order parameter. We shall begin the LDA-self-consistency by using,
to start with, the converged potential parameters from the supercell 
calculations on planar surfaces   
and the equilibrium lattice distances i.e. with a 7.5 $\%$ surface
lattice dilatation. With this starting point the self-consistency
is reached much faster than otherwise.

Figure 2(a) shows the local density of states at a point in the bulk Ag substrate
 (full lines) and that for an Ag atom on the 100 surface of fcc Ag (without
the deposited Fe overlayer) (dotted lines), obtained
by a eight-step recursion process.  We have checked that the recursion does
converge in the sense suggested by Haydock \cite{kn:vol35} and Ghosh {\sl et.al.} \cite{kn:gdm1}
 of the convergence
of integrals of the form 

 \[ \int_{-\infty}^{E} \Phi(E')\; n(E')\; dE' \]

where $\Phi(E)$ is a well-behaved, monotonic function in the integration range.
The Fermi-energy or the chemical potential is calculated from the bulk and is
shown in the Figure 2(a).  As expected we notice that the d-band width decreases
at the surface. This is expected, as the surface atoms are less coordinated
than the bulk (eight on the 100 surface as against twelve in the bulk). There is
also a redistribution of spectral weight in the band. It is clear that the
amount of charge in a Wigner-Seitz sphere around a surface atom is less
than that around a bulk atom. This extra charge leaks out into the so-called
empty-spheres, which carry no atoms but only this leaked charge. By the time
we go down about four layers below the surface, we begin to get local densities
indistinguishable from the bulk results.

Figure 2(b) Shows the local density of states for the up and down electrons in
the Fe overlayer. This is for a perfectly planar overlayer on the 100 surface.
As is usual in either bulk Fe or Fe overlayers on noble metals, the majority
occupied spin band (here up) shows much more structure than the minority occupied one (here
down). Since the Ag d-bands centered round --0.5 ryd do not overlap with either
of the Fe d-bands around --0.2 ryd and --0.1 ryd, there is no significant hybridization
of these two, which usually leads to a widening of the Fe d-bands and consequent
lowering of the local magnetic moment. The Fermi-energy is that of the bulk Ag and
is shown in the figure.

We now alloy the overlayer with empty spheres and re-converge the self-consistent
ASR. In Figure 3 we show the local magnetic moment on a Fe atom in the 
rough overlayer as a function of the Fe concentration in that layer (dotted line) with 7.5 $\%$ surface dilatation. For concentration
x=1 of Fe we obtain the local magnetic moment corresponding to that of Figure 2(b).
The value of 3.17 $\mu_{B}/atom$ is a considerable enhancement on bulk bcc iron local
magnetic moment. The agreement with the supercell calculations is very
close. Bl{\" u}gel has argued \cite{kn:bl} that this can be inferred from
the Stoner criterion because of the narrowing of the overlayer d-bands as compared
with the bulk. As we alloy the overlayer with empty spheres, the local magnetic
moment on an Fe atom increases, until in the extreme case it approaches that of
an isolated Fe atom at $>$ 3.6 $\mu_{B}/atom$. Again we much understand this from
Bl{\" u}gel's argument. We find that the empty spheres hardly inherit any induced
magnetization, as a result as the concentration of empty spheres increase, the
average coordination of Fe atoms decrease, thus increasing the magnetic moment.
In the extreme limit we obtain the case of an Fe impurity atom sitting in a sea
of empty spheres. Its magnetic moment approaches that of a free Fe atom. The
only difference is caused by its hybridization with the Ag substrate. Figure 3
also shows (full lines) the averaged magnetization in the overlayer. This is
defined by : x M$_{Fe}$ + y M$_{ES}$. Since M$_{ES}$ is negligible, this
average overlayer magnetization decreases almost linearly with x and vanishes at x=0. The two
types of magnetization shown in the figure are measured by local magnetic
probes and global magnetization experiments.

Figure 4 shows the local magnetization at atoms in different layers .
We clearly see that there is an induced magnetization in the Ag atoms of the
topmost substrate layers. Magnetization oscillates layer wise into the bulk.

Figure 5 shows the variation of the local magnetic moment at a Fe site (dotted line) and the
averaged magnetic moment in the overlayer as a function of the Warren-Cowley
short-ranged order parameter for (a) x=0.9 and (b) x= 0.75 .
We note that  when the Warren-Cowley parameter indicates phase segregation
 the magnetic moment  shows an increase.
We may understand this behaviour from the following argument : 

 For $\alpha >$ 0 the tendency is towards phase
segregation. Islands of Fe (in our case, clusters of nearest
neighbour atoms)
precipitate in a sea of empty spheres (particularly in the low Fe concentration regime). This situation mimics the islands
and pyramids observed in actual MBE deposited surfaces. A simple
calculation with an isolated five atom nearest neighbour
cluster sitting on the surface shows that the local density of
states on the cluster is much narrower than a homogeneous
distribution of Fe atoms on the surface. This leads to a larger
magnetic moment/atom on the cluster. The maximum  enhancement
of the magnetic moment due to short-ranged clustering is around 3 $\%$. 

Clustering enhancement of magnetic moment competes with the `poisoning'
effect. Interfaces are never sharp, there is always an interdiffusion
of substrate atoms into the surface layer and vice versa. In our final
calculation we have taken a perfectly planar (non rough) monolayer of
Fe on the (100) surface of fcc Ag and allowed upto 10$\%$ interdiffusion
of Fe and Ag atoms in the surface layer and the one just below it. The
surface layer is then an alloy Fe$_{x}$Ag$_{1-x}$ and the next layer
an alloy Ag$_{x}$Fe$_{1-x}$. The following table shows the magnetic moments
in the surface layer for different values of x.

\begin{center}
\begin{tabular}{|c|c|c|c|}\hline
x& Averaged Mag. Mom.& Fe Mag. Mom. & Ag Mag. Mom.  \\ \hline
0.95 & 3.02 & 3.18 & 0.014 \\
0.90 & 2.86 & 3.17 & 0.017 \\ \hline
\end{tabular}
\end{center}
\vskip -0.2cm
\centerline{\small {\bf Table 2} Lowering of Surface magnetism due to `poisoning' by substrate}
\vskip -0.1cm
\centerline{\small All magnetic moments are in ($\mu_{B}/atom$)}

We notice that the depletion of magnetic moment due to poisoning by the substrate
is about 4.5$\%$. In an actual experimental situation both the enhancement effects
due to surface lattice dilatation and clustering and the depletion effect due to poisoning
are present simultaneously. We have a handle on the determination of the lattice
dilatation. Surface roughness may be probed with local techniques like the STM . If
we could determine the amount of interdiffusion, we would be in a position to
quantitatively predict the surface magnetic moment. The conclusion of this communication
is to suggest that the Augmented Space Recursion coupled with any first principles
and accurate technique which
yields a sparse hamiltonian representation (like the TB-LMTO or the screened KKR)
can take into account surface roughness, short-ranged clustering, surface dilatation and interdiffusion effects
accurately and it would be an useful methodology to adopt. 

\vskip 0.5cm
\centerline{\bf ACKNOWLEDGEMENTS}
A.M., G.P.D. and B.S.  should like to thank the ICTP and its Network Project and the DST, India
for financial support of this work. P.B. would like to thank the C.S.I.R., India
for its financial assistance. The collaborative  project between the University
of Warwick and the S.N.Bose National Centre is also gratefully acknowledged. We
should like to thank I. Dasgupta and T. Saha-Dasgupta whose bulk LDA-self-
consistent codes formed the basis of this surface generalization.
\vskip 0.5cm
\centerline{\bf FIGURE CAPTIONS}
\begin{description}
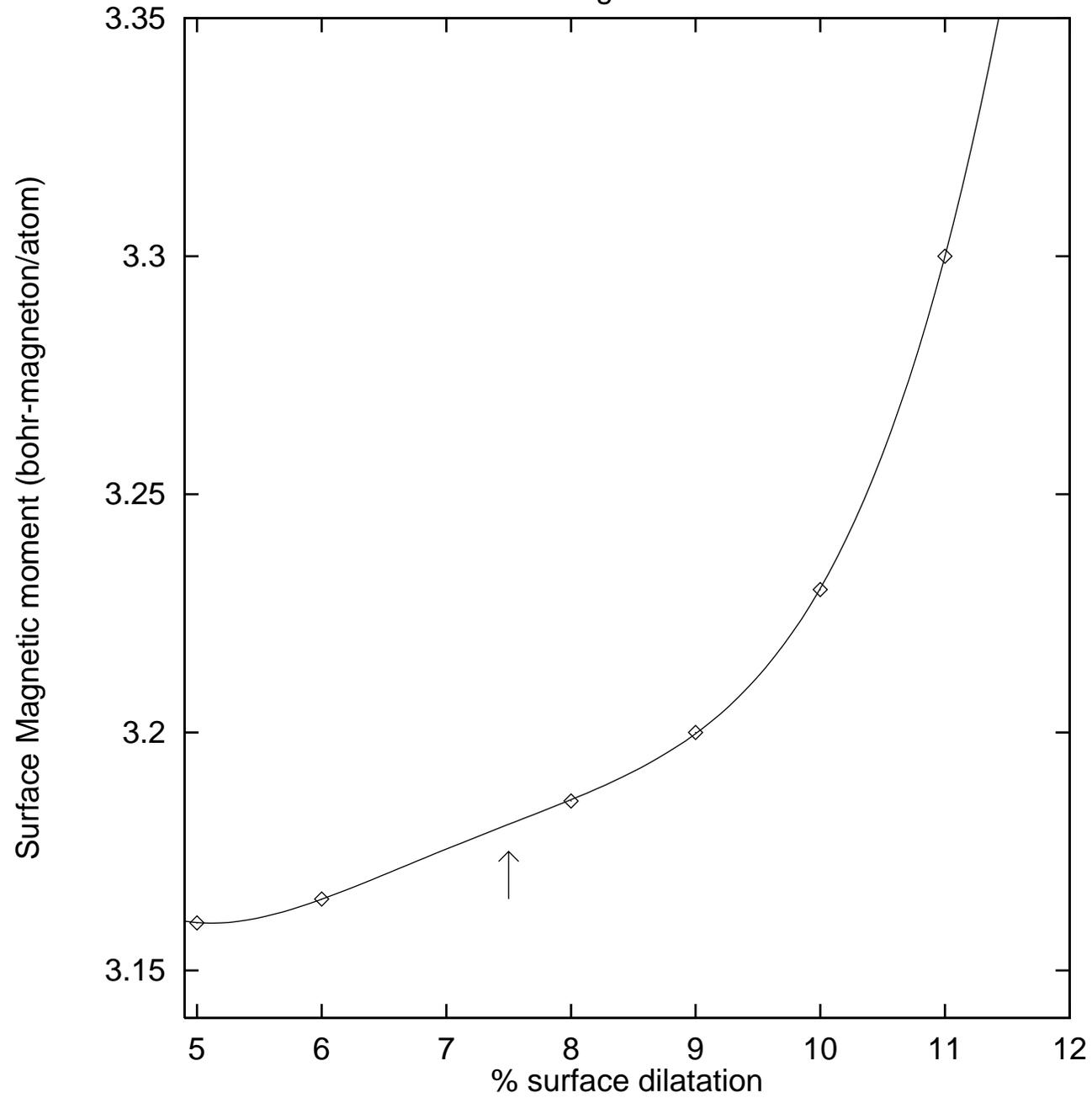
\item[Figure 1] Surface magnetic moment (bohr-magnetons/atom) as a function of
$\%$ surface dilatation (dilatation of the distance between the surface overlayer
and the next layer in the substrate)
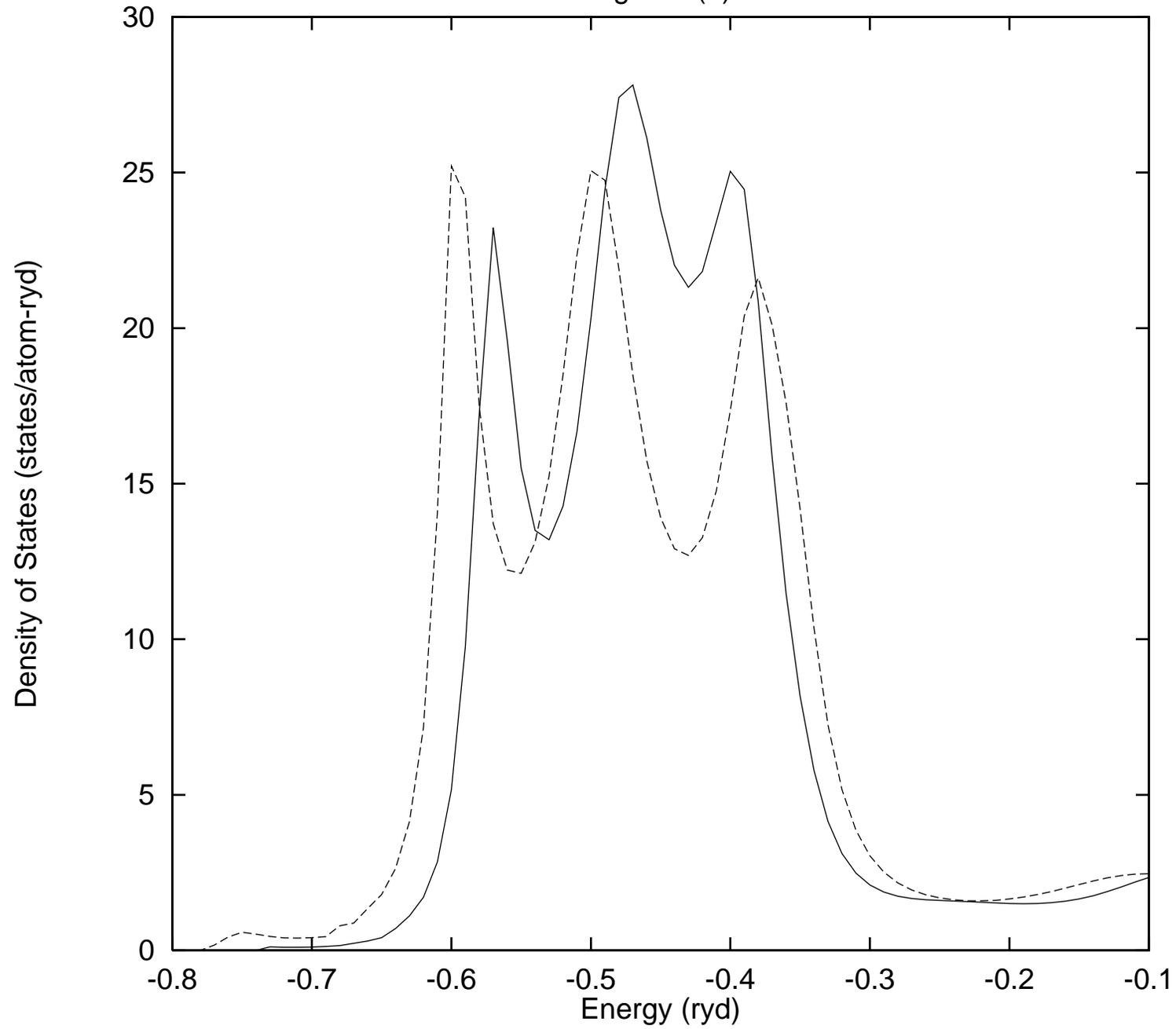
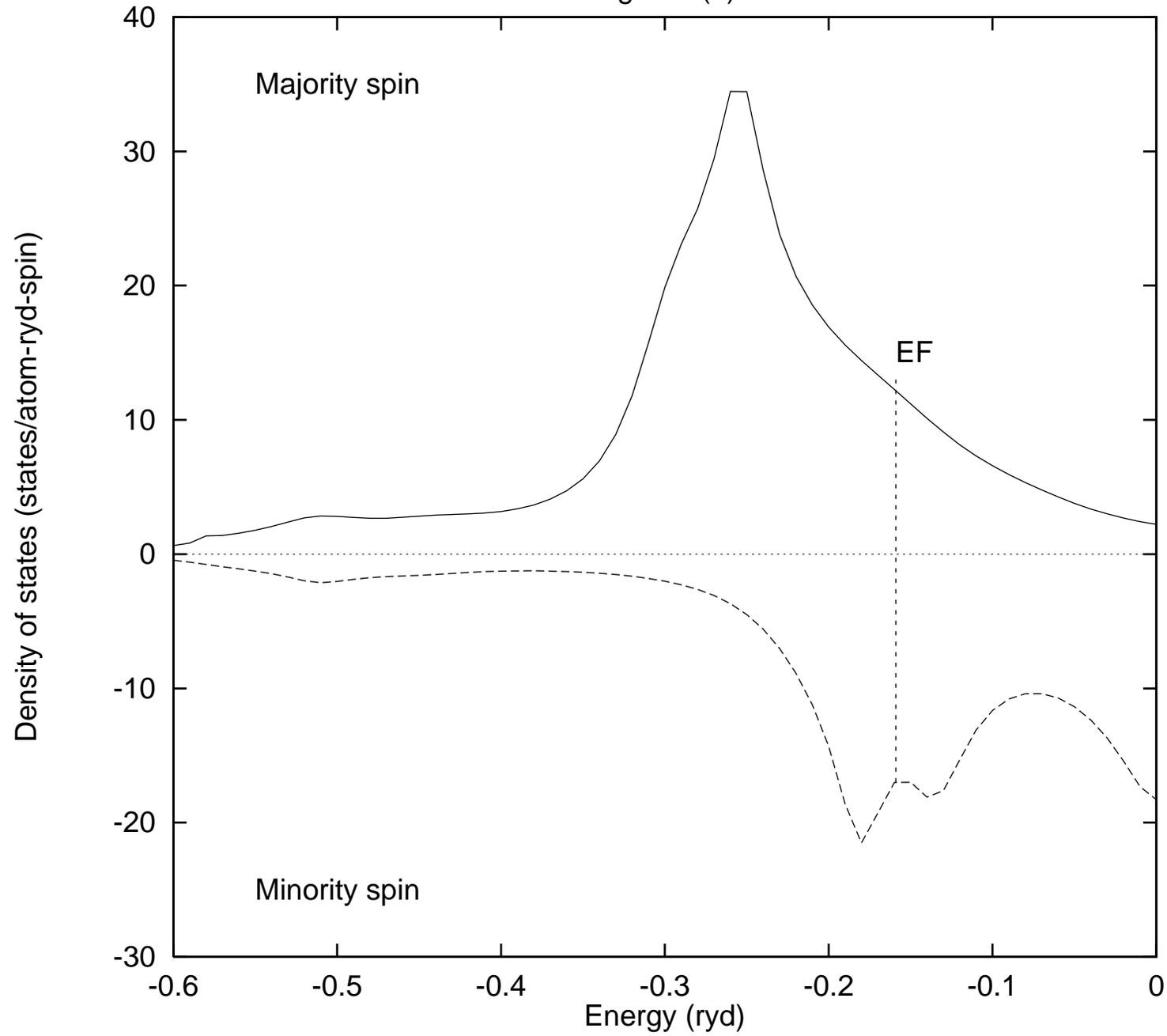
\item [Figure 2]
     \item[(a)] Local density of states at a Ag atom in the bulk (dotted line) and on the (100) surface (full lines)
     \item[(b)] Local density of states at a Fe atom in an overlayer on the (100)
                surface of a Ag substrate. Both the up spin and the down spin densities
                 are shown.
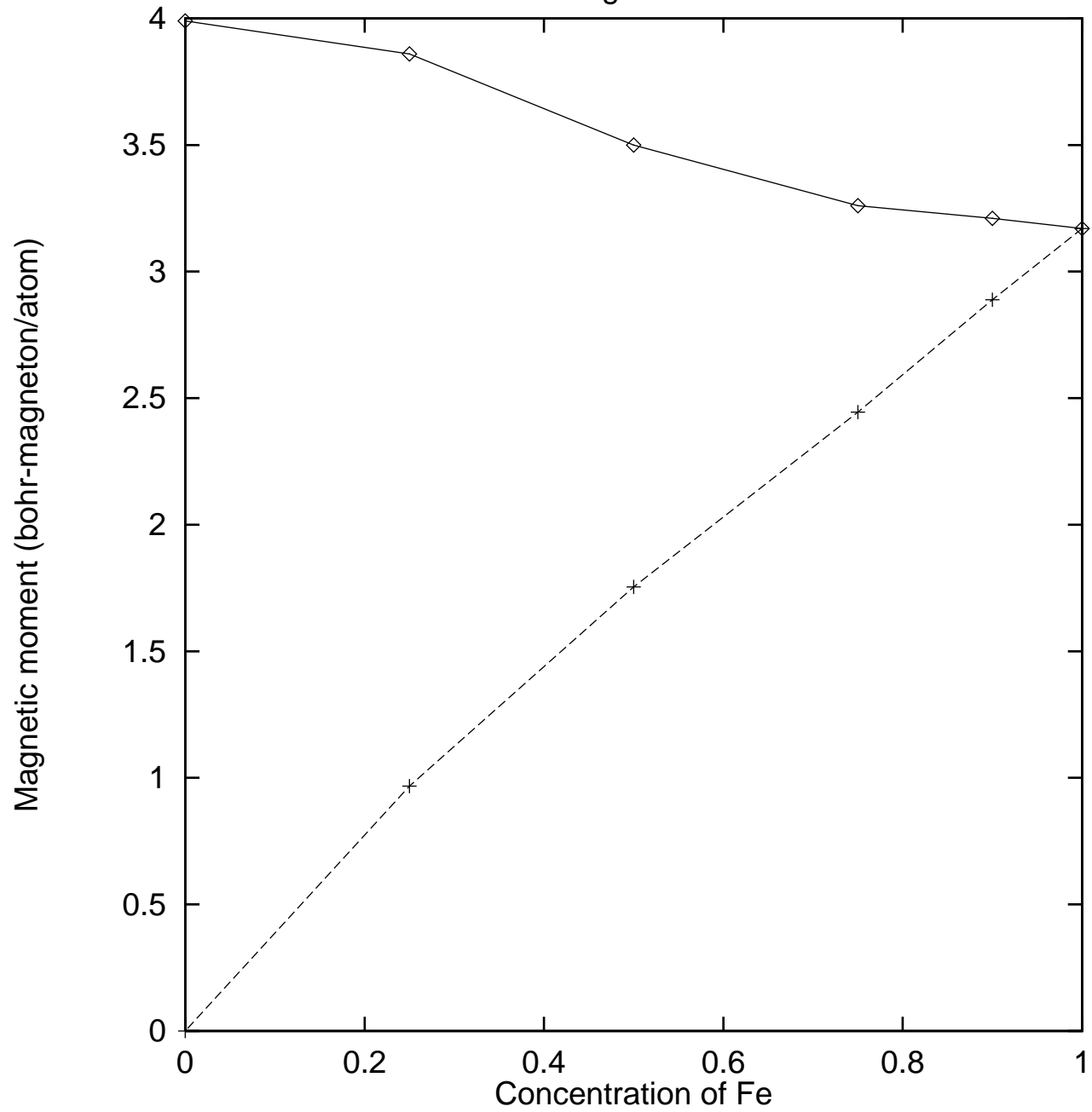
\item[Figure 3] Local magnetic moment (dotted line) and averaged magnetic moment (full line)
                on a Fe atom in a rough overlayer on the (100) surface of a Ag substrate. Roughness
                is modelled by an alloy of Fe and empty spheres. The magnetic moments are shown
                as a function of the concentration of Fe in this model alloy.
Results are for 7.5$\%$ surface dilatation.
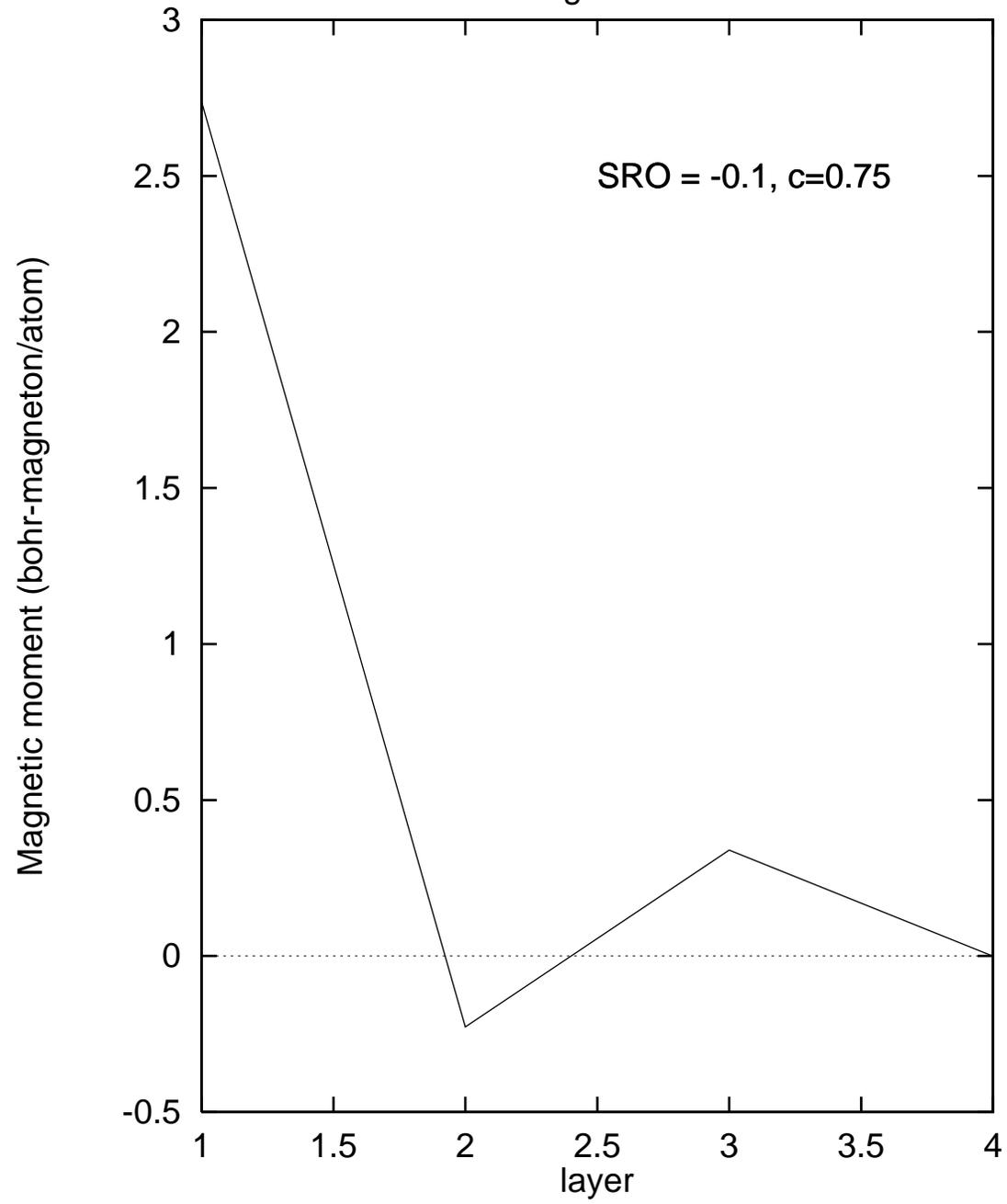
\item [Figure 4] Oscillation of magnetic moment on different layers of a Fe overlayer on the
                (100) surface of a  Ag substrate.
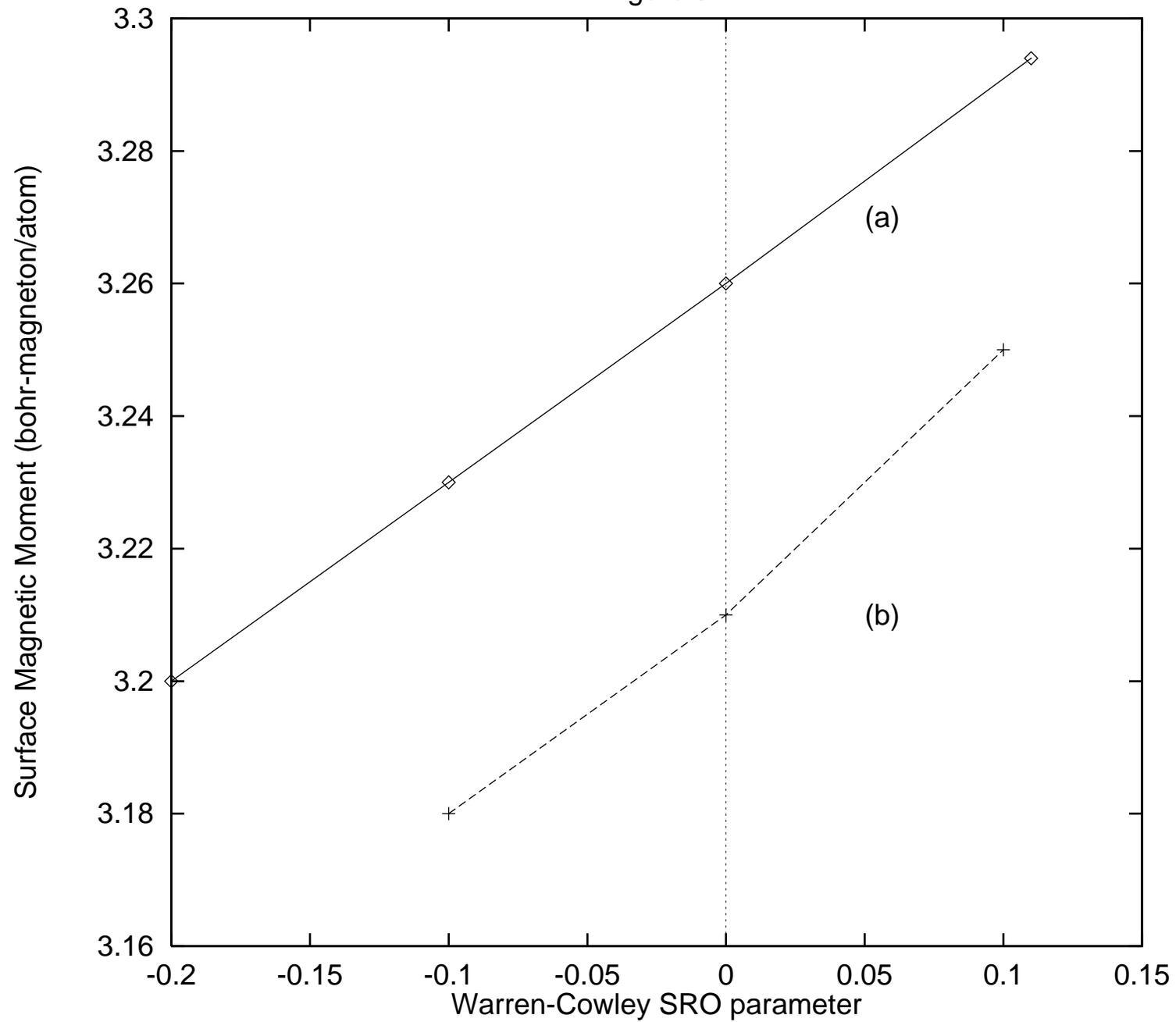
\item[Figure 5] Surface magnetic moment (bohr-magnetons/atom) as a function of
the Warren-Cowley short-ranged order parameter for (a) 90$\%$ Fe 10$\%$ Empty
spheres and (b) 75$\%$ Fe 25$\%$ Empty spheres in the surface overlayer with
7.5$\%$ surface dilatation.
\end{description}

\end{document}